\documentstyle[preprint,aps]{revtex}
\begin{document}
\preprint{SNUTP 95-094}
\title{Is the $ISO(2,1)$ Gauge Gravity equivalent to the Metric Formulation?}
\author{Jin-Ho Cho
\thanks{e-mail address; jhcho@galatica.snu.ac.kr}
}
\address{Center for Theoretical Physics,
Seoul National University, Seoul, 151-742, Korea}
\author{Hyuk-jae Lee
\thanks{e-mail address: hyuklee@buphy.bu.edu}
}
\address{Department of  Physics,
Boston University, Boston MA 02215, USA}
\date{\today}
\maketitle
\begin{abstract}
The quantization of the gravitational Chern-Simons coefficient is
investigated in the framework of $ISO(2,1)$ gauge gravity. Some paradoxes
involved are cured. The resolution is largely based on the
inequivalence of $ISO(2,1)$ gauge gravity and the metric formulation. Both
the Lorentzian scheme and the Euclidean scheme lead to the coefficient
quantization, which means that the induced spin is not quite exotic in this
context.
\end{abstract}
%narrowtext
\section{Introduction}

Despite of its local triviality, (2+1)-dimensional Einstein gravity has been
studied for a long time because it has still rich global structure
\cite{djt} and can be considered as the physics of the
(3+1)-dimensional cosmic strings \cite{vil}. Moreover, the geometry
becomes nontrivial at the semiclassical level \cite{des} where
the gravitational Chern-Simons term (CST),
\begin{eqnarray}
-{1\over{\kappa}}{\cal L}_{CS}
=-{{\alpha}\over{\kappa}}\omega_a\wedge(d\omega^a+{1 \over 3}
\epsilon^a{}_{bc}\omega^b\wedge\omega^c),
\label{cs}
\end{eqnarray}
is generated effectively by the fermion coupled to gravity \cite{pisa}.
Here $\kappa=16\pi G$ and $\alpha$ is a coefficient with the
dimension of length.

One of the most interesting and intricate things of this latter semiclassical
system is the coefficient quantization paradox for the gravitational CST
\cite{jac}.
This problem is well addressed in the papers \cite{per} and can be briefly
summarized as follows.

The subject can be probed in two different
contexts, the dreibein formulation and the metric formulation. In each
context, it can be treated again in two ways, the Lorentzian method
and the Euclidean method. This latter method is based on
the analytic continuation procedure called the Wick rotation, which is
conventionally adopted in most of the quantum field theories.

One important point in the dreibein formulation is to check
whether the gauge symmetry $SO(2,1)$ of Lorentzian system becomes
$SO(3)$ upon the Wick rotation. This is not simple to answer since the
internal gauge symmetry is not affected by such a transformation in the
ordinary Yang-Mills theory. However in the dreibein formulation, 
`Euclideanization' will be shown below to change the gauge symmetry.
Then the third-homotopy argument \cite{des,per} tells us that the
coefficient is quantized in the Euclidean scheme but not in the
Lorentzian method.

Since it is hard to say that those two schemes are not equivalent, this
result gets us into a confusion. Another confusion comes from the
metric formulation, where the local symmetry is known as the
diffeomorphism. Since every diffeomorphism of ${\rm I\!R}^3$ with the given
boundary conditions is homotopic to the identity, the coefficient need not be quantized irrespective
of the metric signature \cite{hai}. This result is in contradiction with that
of gauge formulation since those two formulations are conventionally believed to
be equivalent.

An attempt to solve these paradoxes has been made in \cite{per}. It was shown there that the coefficient need not be quantized in the gauge
formulation even through the Euclidean method. In the proof, $\theta$-sector
term \cite{ish} was introduced. The term has its own
coefficient and has no dynamical relevance due to its topological nature.
Then through the Euclidean argument, one gets the quantization
condition on the {\bf linear combination} of the coefficients of
gravitational CST and $\theta$-sector term rather than on that of the former
term alone. Therefore for any value of the gravitational CS coefficient, one
can adjust the coefficient of $\theta$-sector term to satisfy the
quantization condition. Consequently, one can conclude in all cases that the
gravitational CS coefficient need not be quantized.

However as is said in the paper \cite{per}, the appearance of the
$\theta$-sector term and the physical interpretation of its coefficient
are unclear in the Einstein-Hilbert Lagrangian.
Besides, the new condition tells us that those two coefficients are
mutually dependent. This might be a paradox because there is
no reason to think so in the Lorentzian argument. Another important point to
note is that the dreibein formulation and the metric formulation are {\bf
not} equivalent at least for the above semiclassical system \cite{cho}.
Therefore, we are left with only one paradox: In the dreibein formulation,
the Euclidean scheme and Lorentzian scheme give different results for the
coefficient quantization.

In this paper, we solve this paradox in the context of Poincar\'{e} gauge
gravity. For that purpose, we briefly review in the next section, the
Poincar\'{e} gauge formulation of the (2+1)-dimensional gravity with the
gravitational CST. We discuss about its inequivalence with the metric
formulation, focussing on the resulting geometries a massive point source
makes in each formulation. Section 3 deals with the
Euclidean method to show explicitly that the internal gauge group $ISO(2,1)$
of the Poincar\'{e} gauge gravity changes to $ISO(3)$ upon Wick rotation.
In section 4, it is shown that even the Lorentzian argument results in the
quantization condition for the gravitational CS coefficient.
This amazing result is because the homotopy map in this case
should be further specified to respect the Lorentz structure. Section 5
concludes the paper discussing about the physical implications of our
results and some future prospects on the subject.

\section{(2+1)-dimensional Gravity with the Gravitational CST}

One interesting point in (2+1)-dimension is the existence of another
formulation of gravity, i. e., Poincar\'{e} gauge gravity \cite{wit}.
A characteristic feature of this formulation is that the Einstein-Hilbert
Lagrangian can be obtained just from the Chern-Simons Lagrangian for the
(2+1)-dimensional Poincar\'{e} group, $ISO(2,1)$;
\begin{eqnarray}
-{1\over{\kappa}}{\cal L}_{EH}=-{1\over{\kappa}}<{\cal A}\,,\!\!\!\!\wedge\,
(d{\cal A}+{2 \over 3}{\cal A}\wedge{\cal A})>
=-{1\over{\kappa}}e_a\wedge(2d\omega^a
+\epsilon^a{}_{bc}\omega^b\wedge\omega^c),
\label{1a}
\end{eqnarray}
where the Lie algebra valued one form ${\cal A}=e^a P_a+\omega^a J_a$ is the
gauge connection and the nondegenerate invariant quadratic form $<\,,\,>$ on
the $ISO(2,1)$ group manifold is defined as
$<P_a,J_b>=\eta_{ab},\,<J_a,J_b>=0,\,
<P_a,P_b>=0$. In the above derivation, the following $iso(2,1)$ algebra was
used.
\begin{eqnarray}
[P_a,\,J_b]=\epsilon_{ab}{}^c P_c,\quad
[J_a,\,J_b]=\epsilon_{ab}{}^c J_c,\quad [P_a,\,P_b]=0,
\end{eqnarray}
where $\epsilon^{012}=-\epsilon_{012}=1$. The definition of the torsion and
the curvature are manifest in its field strength components: ${\cal
F}_{\mu\nu}$$\equiv {\cal T}^a_{\ \mu\nu} P_a+{\cal R}^a_{\ \mu\nu} J_a$.

When the matrix composed of the components $e^a{}_\mu$ is invertible, the
Lagrangian (\ref{1a}) is the same as that of the dreibein formulation.
However, this needs not be the case because $e^a{}_\mu$ are the gauge
connection components. To define the invertible
soldering form that geometrically relates the affine tangent space with the
base manifold, we need another important ingredient, the Poincar\'{e}
coordinates, $\phi^a,\,\,(a=0,1,2)$ \cite{cho,grig}. These are nothing but
the isovector components and are concerned with the affine nature
of the group $ISO(2,1)$. We can define gauge invariant metric making use of
the soldering form ${\cal E}^a{}_\mu\equiv{\cal D}_\mu\phi^{a}$. Therefore
in a specific gauge making $\phi^a(x^\mu)=0$, the above Lagrangian
(\ref{1a}) can be considered as that of the dreibein formulation.
Hereafter we use the terminologies, the dreibein formulation and the
Poincar\'{e} gauge formulation interchangeably.

On the other hand, if we restrict the gauge connection components to satisfy
the torsion free condition, ${\cal T}=de+\omega\wedge e=0$, the
same Lagrangian (\ref{1a}) can be considered as that of the metric
formulation. The physics of this
metric formulation is thoroughly studied in the seminal paper \cite{djt}.
For the case with a massive spinning point source, the geometry is locally
trivial but globally has the space-conical and time-helical structure.
As can be noted above, the essential difference between the dreibein
formulation and the metric formulation is concerned with the torsion free
condition. However, despite of this difference, they result in the same
geometry both for the source free case and for the case with a massive
spinning point source \cite{grig}.

The system can be generalized to include the gravitational CST (\ref{cs}).
The term was first introduced by S. Deser, R. Jackiw, and S. Templeton 
in the metric formulation to make the (2+1)-dimensional geometry
locally nontrivial \cite{des}. Afterwards, the term was also shown to be a
possible term generated effectively at the semiclassical level by the
radiative correction of the fermion field coupled to gravity \cite{pisa}.

One can assume the same term to be generated in the
Poincar\'{e} gauge formulation also, although we don't show it explicitly.
This might happen because the spinor transforms under only the subgroup
$SO(2,1)$ and the CST can be generated as in
other (2+1)-dimensional gauge theories \cite{cho2}. Apart from this naive
argument, the term can be formulated in the context of $ISO(2,1)$ gauge
theory by adopting the following generalized quadratic form for the Lie
algebra $iso(2,1)$;
\begin{eqnarray}
<P_a,J_b>=\eta_{ab},\quad<J_a,J_b>=\alpha\eta_{ab},\quad<P_a,P_b>=0.
\label{new}
\end{eqnarray}
This is easily shown to be the most general quadratic form that is
nondegenerate and associative, i. e., has the cyclic property like the trace.
Making use of this quadratic form, one can get both the
Einstein-Hilbert term (\ref{1a}) and the gravitational CST (\ref{cs})
through the CSL \cite{cho}.

One can guess from this enlarged Lagrangian that the geometry
heavily depends on the formulation in consideration. In the metric
formulation one use the torsion free condition. This means the gauge
connection components $\omega^a{}_\mu$ and $e^a{}_\mu$ are not independent
of each other. Indeed the former can be written in terms of the latter as
\begin{eqnarray}
\omega^a{}_{b\mu}=-\partial_{[\mu}e^a{}_{\nu]}e_b{}^\nu
+\partial_{[\mu|}e_{b|\nu]}e^{a\nu}-e_{c\mu}\partial_{[\rho}
e^c{}_{\sigma]}e^{a\rho}e_b{}^\sigma.
\end{eqnarray}
Due to this condition, the derivative order of the gravitational CST,
(\ref{cs}), becomes third while the Einstein-Hilbert term (\ref{1a})
remains to be of the first order. This relative difference in the derivative
order makes the gravitational CST dominant at the large momentum limit while
the Einstein-Hilbert term relevant at the small momentum limit. This also
provides the clue for the topologically massive nature of the graviton.
In the asymptotic region, where the Einstein-Hilbert term becomes dominant,
the geometry becomes that of pure Einstein-Hilbert case except the induced
spin $\alpha m$ \cite{lin}:
\begin{eqnarray}
ds^2\sim-(dt-{{\kappa(\alpha m+\sigma)}\over{4\pi}}d\theta)^2
+{{1}\over{r^{{\kappa m}
\over{2\pi}}}}(dr^2+r^2d\theta^2),
\label{des}
\end{eqnarray}
where $\sigma$ is the particle spin.

On the other hand in the gauge formulation, both terms, (\ref{cs}) and
(\ref{1a}) are of the same derivative order.
Therefore, the dynamics does not depend on the momentum scale at all.
Specifically this case allows the {\bf exact} solution, so 
the above asymptotic solution
extends to all space except the source point. The geometry is
locally flat with no topologically massive graviton even in the
presence of the gravitational CST. This is the result of treating the
connection components $e$ and $\omega$ as independent variables. Moreover,
it opposes the usual belief that the effective Chern-Simons-like terms
assure the topologically massive mode for the gauge
particles. The only effect of the term is the induced spin \cite{cho}.
This is in contrast with the
case of the metric formulation, where it was shown that no exact stationary
solution with the above asymptotically flat limit (\ref{des}) is possible
except the critical case ($\sigma+\alpha m=0$) \cite{cle}.

Another difference is that they choose, in the metric formulation, the
negative sign of Einstein constant to avoid the repulsive force for the
positive mass. Therefore, the deficit angle $8\pi Gm=\kappa m/2$ is negative
and there is no mass bound. However, there is no such problem in the gauge
formulation
because the geometry is locally flat. The positive deficit angle of our case
results in a mass bound $1/4G$. 

This inequivalence is in contrast with the
(3+1)-dimensional case, where the Einstein gravity and Cartan theory are
equivalent in the source free region. In fact, it is a
general feature in the `Palatini' type actions with an extra term,
whose derivative order depends on the Palatini condition.

\section{Euclidean Method}

Now let us go into the main point of this paper, the quantization of
$\alpha$. Making use of the analytic continuation method, we
first work in the Euclidean space, where the variation of the
action under large gauge transformation is easy to visualize as some
topological entity \cite{des}.
In the context of the Poincar\'{e} gauge gravity, we show that the
symmetry $ISO(2,1)$ becomes $ISO(3)$ upon Wick rotation to the Euclidean
space. We also check whether the Lagrangian in the Euclidean space is real
or imaginary. Only the real action can result in the quantization of
$\alpha$ coefficient when it is exponentiated with the extra quantum
coefficient $i/\hbar$ (see Witten in \cite{wit}).

The transfer to the Euclidean space is performed by just changing
Poincar\'{e} coordinates $\phi^a$ and gauge
connections $\omega^a,\,e^a$ to the Euclidean
version. This automatically induces the external spacetime
coordinates transfer through the definition of the metric. The
transfer relations between those two versions are given by
$i\,\phi^0\Rightarrow\tilde{\phi}^3$, $\phi^i\Rightarrow\tilde{\phi}^i$,
$-\omega^0\Rightarrow\tilde {\omega}^3$, $i\,\omega^i\Rightarrow\tilde
{\omega}^i$, $i\,e^0\Rightarrow\tilde{e}^3$ and $e^i\Rightarrow\tilde{e}^i$,
where the tilde `$\sim$' over variables denotes the Euclidean version.
These relations are based on the relations between the two
versions of the soldering form $i{\cal E}^0{}_\mu\Rightarrow
\tilde{\cal E}^3{}_\mu$ and
${\cal E}^1{}_\mu{}\Rightarrow\tilde{\cal E}^i{}_\mu$.

The above transfer relations amount to the change of the internal
gauge symmetry;
\begin{eqnarray}
&&{\cal A}=\omega^aJ_a+e^aP_a=\omega^0J_0+\omega^i J_i+e^0P_0
+e^iP_i\nonumber\\
&&\Rightarrow -\tilde{\omega}^3J_0-i\tilde{\omega}^iJ_i
-i\tilde{e}^3P_0+\tilde{e}^iP_i
=\tilde{{\cal A}},
\end{eqnarray}
where one can read off the Euclidean generators
$\tilde{J}_3=-J_0=J^0$, $\tilde{J}_i=-iJ_i$, $\tilde{P}_3=-iP_0=iP^0$,
$\tilde{P}_i=P_i$ satisfying $iso(3)$ algebra.
This means that the internal Wick rotation of the affine coordinate
$\phi^a$ effectively induces the metric transformation from Lorentzian to
Euclidean and changes the gauge symmetry from $ISO(2,1)$
to $ISO(3)$. This is the crucial difference from the conventional gauge
theory where the transfer of the external spacetime to Euclidean does not
affect the internal space and the gauge group.

According to the above transfer relations, we are led to the
Lagrangian transfer: $i\,{\cal L}_{EH}$$\Rightarrow$
$\tilde{\cal L}_{EH}$, $-{\cal L}_{CS}\Rightarrow\tilde{\cal L}_{CS}$.
Therefore, only CST remains to be real upon the transfer.
Here, one should note that this internal Wick's rotation may change
the topological CSL but does not change the
partition function because the analytic continuation of our concern
is performed in the functional space only.

We next check whether the variation of the Euclidean action under a large
gauge transformation gives some topological invariants concerned with the
third homotopy structure of the transformation group \cite{nas}.
Under the large transformation ${\cal A}\to {\cal A}'=U^{-1}
{\cal A} U +U^{-1}d U$, 
\begin{eqnarray}
\delta\tilde{\cal L} &=&-{1\over 3}<U^{-1}dU\wedge\!\!\!,U^{-1}dU
\wedge U^{-1}dU>_{E}\nonumber\\
&=&{\alpha\over 6}\epsilon_{abc}(\Lambda^{-1}d\Lambda)^a\wedge
(\Lambda^{-1}d\Lambda)^b\wedge(\Lambda^{-1}d\Lambda)^c\nonumber\\
&&+{i\over 2}\epsilon_{abc}(\Lambda^{-1}dq)^a\wedge(\Lambda^{-1}d
\Lambda)^b\wedge(\Lambda^{-1}d\Lambda)^c,
\label{vari}
\end{eqnarray}
where the $<\,,\,>_E$ is the Euclidean inner product and
$\Lambda\in SO(3)$ together with $q\in T(3)$ constitutes the Poincar\'{e}
element $U$.

In the quantum system, the finite action condition requires $U$ tend to
constant in the asymptotic region. This means the field configuration
$U$ can be considered as a map
from $S^3$ to $ISO(2,1)$. In the eq. (\ref{vari}),
only the first term in components can survive upon integration
over $S^3$ since it is renewed as $\alpha Tr\,(\Lambda^{-1}d\Lambda)^3/6$.
This counts the Brouwer degree (winding number) from $S^3$
(the compactified Euclidean space) to $S^3/{\sf Z\!\!Z}^2\sim SO(3)$.
The integration of second term over $S^3$ vanishes because it is
a total divergence. Indeed, one can easily see this making use of the
torsion free equation and the zero curvature equation for the null gauge
distribution, $e=\Lambda^{-1}dq,\,\omega=\Lambda^{-1}d\Lambda$.
Therefore, the invariance of the system under the large
transformation requires the quantization of CST coefficient. In fact,
considering the matter
interaction, we note that extra coefficient $-1/\kappa$
is necessary for the whole gauge field part. Therefore, the
coefficient of CST becomes $-\alpha/\kappa$ which is
to be quantized as $-2\pi\alpha /\hbar\kappa\in{\sf Z\!\!Z}$.
However, the coefficient of the Einstein term, $1/\kappa$ itself has
no reason to be quantized as we saw above.

\section{Lorentzian Method}

The above Euclidean result seems
doubtful because it is in contradiction with the naive Lorentzian
argument that $\pi_3(ISO(2,1))=0$
implies the failure of CST coefficient quantization. To understand this
mystery we work in the Lorentzian spacetime directly. We first ask the
compatibility of the Lorentzian structure on the $S^3$. The answer is
given by the Poincar\'{e}-Hopf theorem; {\it a
connected orientable compact manifold admits a hyperbolic geometry if and
only if its Euler character vanishes} \cite{hop}. Since the Euler
character of $S^3$ vanishes, both the Euclidean structure and the Lorentzian
structure can be constructed on it.

This raises us another question. What is the physical
implication of {\it working in Lorentzian $S^3$ or Euclidean $S^3$}?
On the Lorentzian $S^3$, no timelike curve has end
points \cite{hop}. This might be understood as every timelike curve on $S^3$
should be closed. Indeed, the Lorentz structure on the $S^3$
means that we have a parametrization map from $S^3$ to ${\rm I\!R}$. This
map specifies the value of time parameter for each point of $S^3$. It is
plausible to assume this parametrization map to be continuous otherwise we
cannot define the hypersurface on which the initial physical data are set. 
The assumption forbids those timelike curves to proceed on $S^3$
without overlapping. Therefore for any point of $S^3$, we have a closed
timelike curve passing through it.
One needs not worry about the chronology violation because the parameter
length of the closed timelike curve is very large compared with the ordinary
domain of physics and it is not a relevant thing in our present analysis.

This peculiar spacetime structure can be realized by the Hopf bundle
\cite{nas}, that is, the fibration of $S^3$ into nontrivial (timelike) $S^1$
bundle over (spacelike) $S^2$. We stress here that for
each point on $S^2$,
the time direction along the fiber $S^1$ and any spacelike direction over
$S^2$ cannot be interchanged by any physical process due to the regulation
of $SO(2,1)$. (We see the analogy in ${\rm I\!R^3}$ case, where also both
structures can be installed.
The Lorentzian structure in this case means the foliation of the manifold
${\rm I\!R^3}$ with the family of spacelike slices. This sort of
foliation is also possible for the Euclidean structure. However in the
Lorentzian structure, no physical process
can interchange the line on those two dimensional slices with the line
characterizing the remaining one dimension.)
Physics definitely tells timelikeness
from the spacelikeness. Therefore, given the Lorentzian structure, the
topology $S^3$ should be further specified as the Hopf fibrated $S^3$
where the fiber and the base manifold should be discriminated. This latter
condition is crucial in our argument and we denote this specification as
${}^HS^3$.

Now let us consider the
large gauge transformation of the CS Lagrangian in the Lorentzian
spacetime. The same logic as in the Euclidean argument takes us
to the classification problem of all those transformations (mappings
from ${}^HS^3$ to the $SO(2,1)$ group manifold) according to their homotopy
structure. Here, it is impossible to define the global time over the whole
space $S^2$, that is, one cannot fix a gauge globally for the Hopf bundle
\cite{nas}. Instead, we divide the sphere $S^2$ into two overlapping
regions $S^2_+$
and $S^2_-$ which excludes the South pole and the North pole respectively.
On each region its own global time can be defined and they are connected
with each other in the overlapping region by some $SO(2)$ transformation as
in the Dirac monopole case. This means the whole spacetime topology
${}^HS^3$ is divided into two parts $S^2_+\times S^1$ and $S^2_-\times S^1$.
On the other hand, the topology of $SO(2,1)$ group manifold is two
dimensional hyperboloid $H^2$ fibered with $S^1$ \cite{gil}. 

It is easy to see
the homotopy structure of those mappings from ${}^HS^2$ to $SO(2,1)$ is
nontrivial. Indeed, one can construct a mapping of the
nontrivial Brouwer degree. Since $S^2_\pm$ and $H^2$ are homeomorphic, that
is, topologically equivalent, one can define two one-to-one maps
$\varphi_+: S^2_+\times S^1\rightarrow SO(2,1)$ and
$\varphi_-: S^2_-\times S^1\rightarrow SO(2,1)$ (timelike direction along
$S^1$ is provided with the elements of spacial rotation $SO(2)$ and
the spacelike direction over the regions
$S^2_\pm$ is endowed with the boosting elements of $SO(2,1)$).
It is straightforward to combine those maps to define a new map $\varphi:
{}^HS^2\rightarrow SO(2,1)$. This map is two-to-one and is homotopically
nontrivial, as is obvious from its components maps $\varphi_\pm$.
Consequently, we arrive at the same conclusion as in the Euclidean argument
that the coefficient $\alpha$ should be quantized.

This Minkowski argument is viable for $SU(2)$ gauge
theory also. For that case, one can give one-to-one map from ${}^HS^3$ to
$SU(2)\simeq S^3$ by considering the Hopf fibration of $SU(2)$. This map is
obviously of the Brouwer degree $1$ and it is {\it not} homotopically
equivalent to the constant map (Brouwer degree $0$) as is assured by Hopf
degree theorem \cite{nas}.

\section{Conclusions}

So far, we have discussed about the coefficient quantization for the
gravitational CST and solved some paradoxes involved in this
subject. This resolution is largely based on the inequivalence between the
metric formulation and the dreibein formulation in (2+1)-dimension. Since
those two formulations are different, they need not accord to each other on
the result for the coefficient quantization problem. We also showed it
{\bf wrong} to say that due to the vanishing third homotopy of $ISO(2,1)$,
the coefficient need not be quantized in the Lorentzian scheme of the gauge
formulation. Indeed when we analyze the homotopy structure in
the Lorentzian scheme, we should exclude those homotopy maps that mixes the
timelike curves with the spacelike curves. This restriction gives us further
specified definition of the third homotopy and results in the coefficient
quantization in the Lorentzian scheme. Therefore in the context of gauge
gravity, the coefficient of the gravitational CST should be quantized both
in the Lorentzian scheme and in its analytically continued method, i.e., the
Euclidean scheme. Furthermore as is discussed above, this result does not
contradict with that of the metric formulation, where the coefficient
need not be quantized both in the Lorentzian scheme and in the Euclidean
scheme. Meanwhile we confirmed the naive conjecture that the gauge symmetry
$ISO(2,1)$ should be changed to $ISO(3)$ upon the Wick rotation to the
Euclidean space.

An interesting consequence of this coefficient quantization in the gauge
formulation is that the induced spin is not exotic because its value involves
the gravitational CS coefficient $\alpha$.
This is in contrast with the
inherent spin, which can be fractional in (2+1)-dimension \cite{ger}.
Moreover, the gravitational CST does not any role in the fractional
statistics either. Indeed, it is the Einstein-Hilbert term that dynamically
realizes the spin-statistics theorem through the Aharonov-Bohm effect.
Therefore any kind of spin, whether it is inherent or induced, gives the
same form of Aharonov-Bohm phase upon the interchange of two identical
particles carrying that spin \cite{ort}. The induced spin does not give
rise to the exotic phase since it is not exotic, while the inherent
spin results in the exotic phase. Specifically in the metric formulation,
the induced spin also can be exotic producing anyonic behavior since the
coefficient of the gravitational CST needs not be quantized there
(Deser in \cite{lin}).

Throughout the study, we learned another lesson: It is misconceived that the
CST assures a way of generating mass on the gauge particle. This is possible
only in the case when the dynamical terms of mutually different derivative
order are involved. In the metric formulation of (2+1)-dimensional gravity,
the situation can be met through the torsion free condition.  However in the
dreibein formulation, this condition cannot be satisfied. In this sense, it
is misnomer to call the gravitational CST as the topological mass term. This
terminology makes sense only in the metric formulation.
It would be interesting to study the cosmological consequence of the
difference between those two formulation, on the ground of cosmic string.

The (2+1)-dimensional gravity has been studied in several contexts. J. H.
Horne and E. Witten showed that the topologically massive model of
\cite{des} is equivalent to the Chern-Simons theory for the group $SO(3,2)$
and therefore is finite and exactly soluble \cite{hor}. The equivalence is
in the sense that a solution of the $SO(3,2)$ gauge gravity, with its gauge
specifically fixed so that the connection components $e^a{}_\mu$ form an
invertible matrix, can be a solution of the topologically massive gravity
and vice versa. This means that although those Lagrangians of the two
formulations look mutually different, they result in the same geometry.

In this case, what puzzles us is that the homotopy analysis lead to the
coefficient
quantization for the gravitational CST both in the Lorentzian scheme and
Euclidean scheme. However, one can resolve this problem as follows: In the
analysis of the topologically massive system, one usually use the
diffeomorphism as its symmetry. But as one may note from its sense of the
equivalence with the $SO(3,2)$ gauge gravity, this is quite reduced
symmetry. Indeed, the diffeomorphism, on shell, can be thought of as some
specific form of the conformal transformation \cite{hor} but the reverse is
not true. Therefore, the $SO(3,2)$ gauge gravity can be more than the
topologically massive gravity. We don't know about the true internal symmetry
of this latter system. If it is the diffeomorphism, the coefficient need not
be quantized. If it is a larger symmetry containing the diffeomorphism,
there can be other possibilities.

E. W. Mielke and P. Baekler generalized the topologically massive model by
adding a new translational Chern-Simons term $\sim
\epsilon^{\mu\nu\rho}e^a{}_\mu{\cal T}_{a\nu\rho}$ \cite{mie}. In
collaboration with F. W. Hehl, they further
generalized their model by the cosmological constant term and analyzed its
dynamical structure \cite{bae}. They showed that their model
gives a nontrivial geometry even without the torsion free condition. This
seems to be in contradiction with our derivative order argument. However, in
their generalized topologically massive model, the gauge symmetry is unclear
to us. With a specific set of the coefficients, their model
might be considered rather as (anti-)de Sitter gauge gravity, where the
locally nontrivial ((anti-)de Sitter) structure is assured by the
cosmological constant term (Witten in \cite{wit}).

The coefficient quantization has been also dealt with in several contexts.
M. Henneaux and C. Teitelblim showed that the Dirac magnetic monopole leads
to the quantization of the topological mass even in the abelian CS theory
\cite{hen}. 
Quantization even occurs to the mass: A. Zee showed that the
mass of a particle, around the gravitational analog of Dirac's magnetic
monopole, should be quantized \cite{zee}. Similar quantization was shown to
occur around the spinning cosmic string by P. O. Marzur \cite{mar}.
This seems more likely to happen in the presence of the
gravitational\ CST because the induced spin itself, on which the energy
depends, is quantized. However, this last case
should be further investigated because we do not know the meaning of the CST
in the cosmic string physics \cite{cho2}.
It will be also interesting to probe the cases of the (anti-)de Sitter group.
Since the Poincar\'{e} group can be considered as the vanishing
cosmological constant limit of the (anti-)de Sitter group, such a coefficient
quantization also should occur in the (anti-)de Sitter gauge gravity
\cite{cho3}. 

\section*{Acknowledgement}
One (J.-H.) of the authors is supported by the Korea Science and 
Engineering Foundation in part under the grant number 95-0702-04-01-3
and partially through CTP.


\begin{thebibliography}{9}

\bibitem{djt} S. Deser, R. Jackiw and G. t'Hooft,
             Ann.\ Phys.\ {\bf 152} (1984) 220.
\bibitem{vil} A. Vilenkin,
              Phys.\ Rev.\ {\bf D23} (1981) 852.
\bibitem{des} S. Deser, R. Jackiw and S. Templeton,
             Ann.\ Phys.\ {\bf 140} (1982) 372;
             S. Deser, R. Jackiw and S. Templeton,
             Phys.\ Rev.\ Lett.\ {\bf 48} (1983) 975.
\bibitem{pisa} L. Alvarez-Gaum\'{e}, S. D. Pietra and G. Moore,
             Ann.\ Phys.\ {\bf 163} (1985) 288;
             R. D. Pisarski and S. Rao,
             Phys.\ Rev.\ {\bf D32} (1985) 2081;
             I. Vuorio,
             Phys.\ Lett.\ {\bf B175} (1986) 176.
\bibitem{jac} R. Jackiw, in {\it Relativity, Groups and Topology, Les
              Houches, 1983}, edited by B. de Wit and R. Stora
              (North-Holland, Amsterdam, 1984).
\bibitem{per} R. Percacci,
             Ann.\ Phys.\ {\bf 177} (1987) 27.
\bibitem{hai} A. Hatcher,
             Ann.\ Math.\ {\bf 117} (1983) 533.
\bibitem{ish} C. J. Isham, in {\it Relativity, Groups and Topology, Les
              Houches, 1983}, edited by B. de Wit and R. Stora
              (North-Holland, Amsterdam, 1984).
\bibitem{cho} J.-H. Cho and H.-j. Lee,
              Phys.\ Lett.\ {\bf B351} (1995) 111.
\bibitem{wit} A. Achucarro and P. K. Townsend,
              Phys.\ Lett.\ {\bf B180} (1986) 89;
              E. Witten,
              Nucl.\ Phys.\ {\bf B311} (1988/89) 46.
\bibitem{grig} G. Grignani and G. Nardelli,
               Nucl.\ Phys.\ {\bf B370} (1992) 491.
\bibitem{cho2} J.-H. Cho, (in preparation).
\bibitem{lin} S. Deser,
              Phys.\ Rev.\ Lett.\ {\bf 64} (1990) 611;
              B. Linet,
              Gen.\ Rel.\ Grav.\ {\bf 23} (1991) 15.
\bibitem{cle} G. Cl\'{e}ment,
              Class.\ Quantum \ Grav.\ {\bf 7} (1990) L193.
\bibitem{nas} C. Nash and S. Sen, {\it Topology and Geometry for
              Physicists} (Academic Press, Inc., London, 1983);
              V. Guillemin and A. Pollack, {\it Differential
              Topology} (Prentice-Hall, Inc., Englewood Cliffs,
              New Jersey, 1974).
\bibitem{hop} H. Hopf,
              Math.\ Ann.\ {\bf 96} (1926) 225;
              R. P. Geroch,
              J.\ Math.\ Phys.\ {\bf 8} (1967) 782.
\bibitem{gil} R. Gilmore, {\it Lie Groups, Lie Algebras, and Some
              of Their Applications} (John Wiley \& Sons, New York, 1974).
\bibitem{ger} P. de S. Gerbert,
              Nucl.\ Phys.\ {\bf B346} (1990) 440.
\bibitem{ort} M. E. Ortiz,
              Nucl.\ Phys.\ {\bf B363} (1991) 85.
\bibitem{hor} J. H. Horne and E. Witten,
              Phys.\ Rev.\ Lett.\ {\bf 62} (1989) 501.
\bibitem{mie} E. W. Mielke and P. Baekler,
              Phys.\ Lett.\ {\bf A156} (1991) 399.
\bibitem{bae} P. Baekler, E. W. Mielke and F. W. Hehl,
              Il\ Nuovo\ Cim.\ {\bf 107B} (1992) 91.
\bibitem{hen} M. Henneaux and C. Teitelboim,
              Phys.\ Rev.\ Lett.\ {\bf 56} (1986) 689.
\bibitem{zee} A. Zee,
              Phys.\ Rev.\ Lett.\ {\bf 55} (1985) 2379.
\bibitem{mar} P. O. Marzur,
              Phys.\ Rev.\ Lett.\ {\bf 57} (1986) 929.
\bibitem{cho3} J.-H. Cho and H.-j. Lee, preprint SNUTP 95-058.

\end{thebibliography}
\end{document}